\documentclass[11pt]{article}
\usepackage{exscale}\usepackage{graphicx}
\def\d{\mbox{\rm d}}
\begin{document}
\title{Lagrangians for biological models}
 \author{M.C. Nucci  $\;$ and $\;$ K.M. Tamizhmani\footnote{Permanent address: Department of
Mathematics, Pondicherry University, R. V. Nagar, Kalapet,
Pondicherry, 605 014,  India}}
\date{Dipartimento di Matematica
e Informatica, Universit\`a di Perugia \&  INFN-Perugia, 06123 Perugia, Italy}

 \maketitle
 \begin{abstract}
We show that a method presented in [S.L. Trubatch and A. Franco,
Canonical Procedures for Population Dynamics, J. Theor. Biol. 48
(1974), 299-324] and later in [G.H. Paine, The development
of Lagrangians for biological models, Bull. Math. Biol. 44
(1982) 749-760] for finding Lagrangians
of classic models in biology, is actually based on finding the
Jacobi Last Multiplier of such models. Using known properties of
Jacobi Last Multiplier we show how to obtain linear Lagrangians of
those first-order systems and nonlinear Lagrangian of the corresponding
single second-order equations that can be derived from them, even in the case
where those authors failed such as the
 host-parasite model.
 \end{abstract}
PACS: 02.30.Hq, 02.30.Xx, 45.20.Jj, 87.23.Cc\\
Keywords: Jacobi Last Multiplier, Lagrangian, Population dynamics
\section{Introduction}
About 36 years ago, Trubatch and Franco published a paper
\cite{TruF} in which they presented an explicit algorithm for
constructing Lagrangians of some biological systems, namely the
classical Volterra-Lotka's model \cite{Volterra}, the Gompertz's
model \cite{Gompertz}, the Verhulst's model \cite{Verh}, and an
host-parasite model \cite{LesG}.

Their method for finding a Lagrangian of a second-order equation
is, as they state, that by Havas \cite{Havas 57} who based his
method on Helmholtz's work \cite{Helm 87}. Neither Helmholtz nor
Havas ever acknowledged the use of the Jacobi Last Multiplier
 in order to find Lagrangians of a second-order equation
\cite{JacobiVD}\footnote{An English translation is now available \cite{VDEng}.}, \cite{Whittaker}.

Indeed, the method by Trubatch and Franco is based on finding a
function $f$ that satisfies their equation (6) and is nothing else
than the Jacobi Last Multiplier. Because they did not know the
properties of the Jacobi Last Multiplier they were unable to find
 a Lagrangian for the
host-parasite model. In fact they found just a linear Lagrangian
of this model and stated  explicitly ``In general, there is no
relation between the linear Lagrangians of this section and the
non-linear ones of the previous section for the same model
systems." In this paper we prove that they were wrong.

It is interesting to note that the method by Trubatch and Franco
for finding linear Lagrangians is that introduced by Kerner
\cite{Kerner 71}. Again Trubatch and Franco did not realize that
their key-function $W$ that satisfies their equation (50a) is
nothing else than the Jacobi Last Multiplier\footnote{Actually
Havas and Kerner never acknowledged each other work, although, at
least once, they were presenting at the same meeting in the same
section \cite{HavasB} and \cite{KernerB}.}.

Eight years later, Paine \cite{Paine} published a paper on the
same subject and based his work on the method introduced by Kerner
\cite{Kerner 71}, and cited Helmholtz's work \cite{Helm 87} as
well. Of course,  the method proposed by Paine is based on a function
$g$ that is actually the Jacobi Last Multiplier of the
two-dimensional systems that he studies.

Paine posed the following questions: ``What are the criteria that
a system of ordinary differential equations must satisfy to assure
the existence of a Lagrangian?" (omissis) ``Does there exist an
algorithm that enables one to construct the Lagrangian from the
dynamical equations?"

Strange enough, he did not mention  the previous work by Trubatch
and Franco \cite{TruF}. Paine's examples are the Volterra-Lotka's
model similar to that studied by Trubatch and Franco \cite{TruF},
and two quite trivial linear systems.

In this paper we show that recognizing that the key-function for
finding a Lagrangian is the Jacobi Last Multiplier permits to
obtain all the results in \cite{TruF} in a simple and complete
 way and furthermore where Trubatch and Franco fail, namely the
model of host-parasite, Jacobi Last Multiplier prevails by
yielding a suitable Lagrangian.

The paper is organized in the following way. In the next section
we  recall the properties of the Jacobi Last Multiplier, its
well-known connection with Lagrangians of  second-order equations \cite{Whittaker},
and the connection with the Lagrangian of systems of two first order equations:
Noether's theorem \cite{Noether} is also presented. In
section 3 we apply the method of the Jacobi Last Multiplier to the
same systems as given in \cite{TruF}, and their equivalent single second-order equation.
 The last section contains
some final remarks.

\section{The method by Jacobi}
The method of the Jacobi Last Multiplier (\cite{Jacobi 44 a},
\cite{Jacobi 44 b}, \cite{Jacobi 45}, \cite{JacobiVD}) provides a
means to determine all the solutions of the partial differential
equation
\begin {equation}
\mathcal{A}f = \sum_{i = 1} ^n a_i(x_1,\dots,x_n)\frac {\partial
f} {\partial x_i} = 0 \label {2.1}
\end {equation}
or its equivalent associated Lagrange's system
\begin {equation}
\frac {\d x_1} {a_1} = \frac {\d x_2} {a_2} = \ldots = \frac {\d
x_n} {a_n}.\label {2.2}
\end {equation}
In fact, if one knows the Jacobi Last Multiplier and all but one
of the solutions, namely $n-2$ solutions, then the last solution
can be obtained by a quadrature. The Jacobi Last Multiplier $M$ is
given by
\begin {equation}
\frac {\partial (f,\omega_1,\omega_2,\ldots,\omega_{n- 1})}
{\partial (x_1,x_2,\ldots,x_n)}
 = M\mathcal{A}f, \label {2.3}
\end {equation}
where
\begin {equation}
\frac {\partial (f,\omega_1,\omega_2,\ldots,\omega_{n- 1})}
{\partial (x_1,x_2,\ldots,x_n)} = \mbox {\rm det}\left [
\begin {array} {ccc}
\displaystyle {\frac {\partial f} {\partial x_1}} &\cdots &\displaystyle {\frac {\partial f} {\partial x_n}}\\
\displaystyle {\frac {\partial\omega_1} {\partial x_1}} & &\displaystyle {\frac {\partial\omega_1} {\partial x_n}}\\
\vdots & &\vdots\\
\displaystyle {\frac {\partial\omega_{n- 1}} {\partial x_1}}
&\cdots &\displaystyle {\frac {\partial\omega_{n- 1}} {\partial
x_n}}
\end {array}\right] = 0 \label {2.4}
\end {equation}
and $\omega_1,\ldots,\omega_{n- 1} $ are $n- 1 $  solutions of
(\ref {2.1}) or, equivalently, first integrals of (\ref {2.2})
independent of each other. This means that  $M$ is a function of
the variables $(x_1,\ldots,x_n)$ and  depends on the chosen $n-1$
solutions, in the sense that it varies as they vary. The essential
properties of the Jacobi Last Multiplier are:
\begin{description}
\item{ (a)} If one selects a different set of $n-1$ independent
solutions $\eta_1,\ldots,\eta_{n-1}$ of equation (\ref {2.1}),
then the corresponding last multiplier $N$ is linked to $M$ by the
relationship:
$$
N=M\frac{\partial(\eta_1,\ldots,\eta_{n-1})}{\partial(\omega_1,
\ldots,\omega_{n-1})}.
$$
\item{ (b)} Given a non-singular transformation of variables
$$
\tau:\quad(x_1,x_2,\ldots,x_n)\longrightarrow(x'_1,x'_2,\ldots,x'_n),
$$
\noindent then the last multiplier $M'$ of  $\mathcal{A'}F=0$ is
given by:
$$
M'=M\frac{\partial(x_1,x_2,\ldots,x_n)}{\partial(x'_1,x'_2,\ldots,x'_n)},
$$
where $M$ obviously comes from the $n-1$ solutions of
$\mathcal{A}F=0$ which correspond to those chosen for
$\mathcal{A'}F=0$ through the inverse transformation $\tau^{-1}$.
\item{ (c) } One can prove that each multiplier $M$ is a solution
of the following
 linear partial differential equation:
 \begin {equation}
\sum_{i = 1} ^n \frac {\partial (Ma_i)} {\partial x_i} =
0;\label{Meqi} \end {equation}
\noindent viceversa every solution
$M$ of this equation is a Jacobi Last Multiplier.
\item{ (d) } If one
knows two Jacobi Last Multipliers $M_1$ and $M_2$ of equation
(\ref {2.1}), then their ratio is a solution $\omega$ of (\ref
{2.1}), or, equivalently,  a first integral of (\ref {2.2}).
Naturally the ratio may be quite trivial, namely a constant.
Viceversa the product of a multiplier $M_1$ times any solution
$\omega$ yields another last multiplier
$M_2=M_1\omega$.\end{description}

There is an obvious corollary to the results of Jacobi mentioned
above. In the case that there exists a constant multiplier, then
any other Jacobi Last Multiplier is a first integral.

Another property of the Jacobi Last Multiplier is  its (almost
forgotten) relationship with the Lagrangian, $L=L(t,x,\dot x)$,
for any second-order equation
\begin{equation}
\ddot x=\phi(t,x,\dot x) \label{geno2}
\end{equation}
i.e. \cite{JacobiVD} (Lecture 10), \cite{Whittaker}
\begin{equation}
M=\frac{\partial^2 L}{\partial \dot x^2} \label{relMLo2}
\end{equation}
where $M=M(t,x,\dot x)$ satisfies the following equation
\begin{equation} \frac{{\rm d}}{{\rm d} t}(\log M)+\frac{\partial \phi}{\partial
\dot x} =0.\label{Meq}
\end{equation}
Then equation (\ref{geno2}) becomes the Euler-Lagrangian equation:
\begin{equation}
-\frac{{\rm d}}{{\rm d} t}\left(\frac{\partial L}{\partial \dot
x}\right)+\frac{\partial L}{\partial x}=0. \label{ELo2}
\end{equation}
The proof is given by taking the derivative of (\ref{ELo2}) by
$\dot x$ and showing that this yields (\ref{Meq}).
 If one knows a Jacobi Last Multiplier, then $L$ can be
  obtained by a double integration, i.e.:
\begin{equation}
L=\int\left (\int M\, {\rm d} \dot x\right)\, {\rm d} \dot
x+\ell_1(t,x)\dot x+\ell_2(t,x), \label{lagrint}
\end{equation}
where $\ell_1$ and $\ell_2$ are functions of $t$ and $x$ which
have to satisfy a single partial differential equation related to
(\ref{geno2}) \cite{laggal}. As it was shown in \cite{laggal},
$\ell_1, \ell_2$ are related to the gauge function $F=F(t,x)$. In
fact, we may assume
\begin{eqnarray}
\ell_1&=&  \frac{\partial F}{\partial x}\nonumber\\
\ell_2&=& \frac{\partial F}{\partial t} +\ell_3(t,x)
\label{gf1f2o2}
\end{eqnarray}
where $\ell_3$ has to satisfy the mentioned partial differential
equation and $F$ is obviously arbitrary.

In \cite{TruF} it was shown that a system of two first-order
ordinary differential equations
\begin{eqnarray}
\dot u_1&=&\phi_1(t,u_1,u_2) \nonumber \\ \dot
u_2&=&\phi_2(t,u_1,u_2) \label{sys}
\end{eqnarray}
always admits a linear Lagrangian of the form
\begin {equation}
L=U_1(t,u_1,u_2)\dot u_1+U_2(t,u_1,u_2)\dot u_2 -V(t,u_1,u_2).
\label{LagsysTB}
\end{equation}
The key is a function $W$ such that\footnote{In \cite{TruF} this
formula contains an inessential multiplicative constant, namely
the integer 2.}
\begin {equation}
W=-\frac{{\rm \partial}U_1}{{\rm \partial}u_2} =\frac{{\rm
\partial}U_2}{{\rm \partial}u_1}  \label{condW}
\end{equation}
and
\begin {equation}
\frac{{\rm d}}{{\rm d} t}(\log W)+\frac{\partial \phi_1}{\partial
u_1}+\frac{\partial \phi_2}{\partial u_2} =0.\label{Weq}
\end{equation}
It is obvious that equation (\ref{Weq}) is the equation
(\ref{Meqi}) of the Jacobi Last Multiplier for system (\ref{sys}).
Therefore once a Jacobi Last Multiplier $M(t,u_1,u_2)$ has been
found, then a Lagrangian of system (\ref{sys}) can be obtained by
two integrations, i.e.:
\begin {equation}
L=\left(\int M\,{\rm d} u_1 \right)\dot u_2-\left(\int M\,{\rm d}
u_2 \right)\dot u_1 +g(t,u_1,u_2)+\frac{{\rm d}}{{\rm
d}t}G(t,u_1,u_2), \label{Lagsys}
\end{equation}
where $g(t,u_1,u_2)$ satisfies two linear differential equations
of first order that can be always integrated, and $G(t,u_1,u_2)$
is the arbitrary gauge function\footnote{The gauge function was
not taken into consideration in \cite{TruF}.} that should be taken
into consideration in order to correctly apply Noether's theorem
\cite{Noether}. If a Noether's symmetry
\begin{equation}
\Gamma=\xi(t,u_1,u_2)\partial_t+\eta_1(t,u_1,u_2)\partial_{u_1}+\eta_2(t,u_1,u_2)\partial_{u_2}
\end{equation}
exists for the Lagrangian $L$ in (\ref{Lagsys}) then a first
integral of system (\ref{sys}) is
\begin{equation}
-\xi L-\frac{\partial L}{\partial \dot u_1}(\eta_1-\xi \dot u_1)
-\frac{\partial L}{\partial \dot u_2}(\eta_2-\xi \dot
u_2)+G(t,u_1,u_2). \label{intsys}
\end{equation}
We underline that  $\dot u_1$ and $\dot u_2$  always disappear
from the expression of the first integral (\ref{intsys}) thanks to the linearity
of the Lagrangian (\ref{Lagsys}) and formula (\ref{Lagsys}).

\section{Some biological examples from \cite{TruF}}
\subsection{ Volterra-Lotka's model}
The Volterra-Lotka's model considered in \cite{TruF} is the
following:
\begin{eqnarray} \dot w_1&=& w_1(a + b w_2) \nonumber\\
 \dot w_2 &=& w_2(A + B w_1) \label{VL}.
\end{eqnarray}
In order to simplify system (\ref{VL}) we follow \cite{TruF} and
introduce the change of variables
\begin{equation}
w_1=\exp(r_1),\quad\quad w_2=\exp(r_2)\label{VLtr}
\end{equation}
and then system (\ref{VL})  becomes
\begin{eqnarray} \dot r_1&=& b\exp(r_2) + a \nonumber\\
 \dot r_2 &=& B\exp(r_1) + A\label{VLt}.
\end{eqnarray}
An obvious Jacobi Last Multiplier of this system is a constant,
say 1, and consequently by means of (\ref{Lagsys}) a linear
Lagrangian of system (\ref{VLt}) is
\begin{equation}
L_{[r]}=r_1\dot r_2-r_2\dot r_1+2( - B\exp(r_1) + b\exp(r_2) -
Ar_1 + ar_2)+\frac{{\rm d}}{{\rm d}t} G(t,r_1,r_2)
\end{equation}
which (minus the gauge function $G$) was found in \cite{TruF}.
Moreover we can    derive a Jacobi Last Multiplier for the
Volterra-Lotka system (\ref{VL}) by using property (b). In fact we
have to calculate the Jacobian of the transformation (\ref{VLtr})
between $(w_1,w_2)$ and $(r_1,r_2)$ and this yields a Jacobi Last
Multiplier of system (\ref{VL}), i.e.
\begin {equation}
M_{[w]} =M_{[r]}\frac{\partial(r_1,r_2)}{\partial(w_1,w_2)}=\left
|
\begin {array} {cc}
{\displaystyle{\frac{1}{w_1}}}& 0\\ [0.3cm] 0 &
{\displaystyle{\frac{1}{w_2}}}
\end {array}\right|=\frac{1}{w_1w_2}. \label {VLM}
\end {equation}
Finally, formula (\ref{Lagsys})  yields a linear Lagrangian of
system (\ref{VL})
\begin{equation}
L_{[w]}=\log(w_1)\frac{\dot w_2}{w_2}-\log(w_2)\frac{\dot
w_1}{w_1}+2( - A\log(w_1) + a\log(w_2) - Bw_1 + bw_2) +\frac{{\rm d}}{{\rm d}t}
G(t,w_1,w_2). \label{LwVL}
\end{equation}
This Lagrangian was not obtained in \cite{TruF}. We note that
(\ref{VL}) is autonomous and therefore invariant under time
translation, namely  $\partial_t$. It is easy to show that the
Lagrangian $L_{[w]}$ in (\ref{LwVL}) yields a time-invariant first
integral through Noether's theorem \cite{Noether}, i.e.:
\begin{equation}
-L_{[w]}+\dot w_1\frac{\partial L_{[w]}}{\partial \dot w_1} +\dot
w_2\frac{\partial L_{[w]}}{\partial \dot w_2}=A\log(w_1) -
a\log(w_2) + Bw_1 - bw_2=const
\end{equation}

Following \cite{TruF} we can transform  system (\ref{VLt}) into an
equivalent second-order ordinary differential equation by
eliminating, say, $r_1$. In fact from the second equation in
(\ref{VLt}) one gets
\begin {equation}r_1=\log\left(\frac{\dot r_2 - A}{B}\right),\label{r1}\end{equation}
and the equivalent second-order
equation in $r_2$ is the following
\begin {equation}
\ddot r_2=- \Big(b\exp(r_2)+a\Big)(A- \dot r_2). \label{VLr2}
\end {equation}
A Jacobi Last Multiplier for this equation has to satisfy equation
(\ref{Meq}), i.e.:
\begin {equation}
\frac{{\rm d}}{{\rm d} t}(\log M)+ b\exp(r_2) + a =0\end{equation}
namely
\begin {equation} \frac{{\rm d}}{{\rm d} t}(\log M)+ \dot r_1=0,
\end{equation}
by taking into account the first equation in (\ref{VLt}), and
consequently we get the following Jacobi Last Multiplier for
equation (\ref{VLr2}):
\begin {equation} M_1=\exp(-r_1)=\frac{B}{\dot r_2-A}, \label{VLM1}
\end{equation}
the last equality thanks to (\ref{r1}). Then a Lagrangian  can be
  obtained by a double integration as in (\ref{lagrint}),
i.e.
\begin {equation}
L_1=B\Big(( \dot r_2-A) \log(A -\dot r_2)  -  \dot r_2+b\exp(r_2)
+ ar_2\Big)+\frac{{\rm d}}{{\rm d}t} F(t,r_2). \label{VLL1}
\end{equation}
The same Lagrangian (minus the gauge function $F$) was obtained in
\cite{TruF}. In order to show the power of the Jacobi's method we
derive at least another Lagrangian for equation (\ref{VLr2}).

We note that (\ref{VLr2}) is autonomous and therefore invariant
under time translation. It is easy to show that the Lagrangian
$L_1$ in  (\ref{VLL1}) yields a time-invariant first integral,
 through Noether's theorem \cite{Noether}, i.e.:
\begin{equation}
I_1=-a r_2 + \dot r_2 + A\log(A- \dot r_2)- b\exp(r_2)=const
\label{VLint}
\end{equation}
 As a consequence of the  property (d) of the Jacobi last multiplier,
the product of a Jacobi last multiplier $M_1$ as in (\ref{VLM1})
and a first integral  $I_1$ as in (\ref{VLint}) of equation
(\ref{VLr2}) yields another Jacobi last multiplier, i.e.
\begin {equation}
M_2=M_1 I_1 = \frac{B}{A-\dot r_2}\Big(a r_2 - \dot r_2 - A\log(A-
\dot r_2)+ b\exp(r_2)\Big)\label{VLM2}
\end{equation}
and therefore we can obtain a second Lagrangian of equation
(\ref{VLr2}), i.e.
\begin {eqnarray}
L_2&=&-\frac{B}{2} \Big(  \left(A\log(A - \dot r_2) -
2ar_2\right)(A - \dot r_2)\log(A - \dot r_2) \nonumber\\&&- (2ar_2
+ \dot r_2)
     \dot r_2 - 2b\exp(r_2)\left((A - \dot r_2)\log(A - \dot r_2) + \dot r_2\right)
  \nonumber\\&&   + b^2\exp(2r_2)
     + 2abr_2\exp(r_2) +a^2r_2^2 \Big)+\frac{{\rm d}}{{\rm d}t} F(t,r_2)
\label{VLL2}
\end{eqnarray}
This Lagrangian yields another time invariant first integral which
is just the square of $I_1$
in (\ref{VLint}).\\
We can keep using property (d) to derive more and more Jacobi last
multipliers and therefore Lagrangians of equation (\ref{VLr2}). In
fact other Jacobi last multipliers can be obtained by simply
taking any function of the first integral $I_1$ in (\ref{VLint})
and multiplying it for either $M_1$ in (\ref{VLM1}) or $M_2$ in
(\ref{VLM2}), and so on ad libitum.

\subsection{Gompertz's model}

The Gompertz's model considered in \cite{TruF} is the
following\footnote{In \cite{TruF} some parentheses are missing: an
obvious missprint.}:
\begin{eqnarray} \dot w_1&=& w_1 \left(A \log\left(\frac{w_1}{m_1}\right) + B w_2\right)\nonumber\\
 \dot w_2 &=&  w_2\left( a \log\left(\frac{w_2}{m_2}\right) + b w_1\right).  \label{Gom}
\end{eqnarray}
In order to simplify system (\ref{Gom}) we follow \cite{TruF} and
introduce the change of variables
\begin{equation}
w_1=m_1\exp(r_1),\quad\quad w_2=m_2\exp(r_2)\label{Gomtr}
\end{equation}
and then system (\ref{Gom})  becomes
\begin{eqnarray} \dot r_1&=& m_2 B\exp(r_2) + Ar_1\nonumber\\
 \dot r_2 &=& m_1 b\exp(r_1) + a r_2  \label{Gomt}.
\end{eqnarray}
It is easy to derive a Jacobi Last Multiplier for this system from
(\ref{Meqi}), i.e.
\begin{equation}
\frac{{\rm d}}{{\rm d}t}\log
\left(M_{[r]}\right)=-(a+A)\Longrightarrow M_{[r]}=
\exp[-(a+A)t]\label{GomMr}
\end{equation}
and therefore the following Lagrangian
\begin{equation}
L_{[r]}=\exp[-(a+A)t][r_1\dot r_2-r_2\dot r_1- 2m_1b\exp(r_1) +
2m_2B\exp(r_2) + (A-a)r_1r_2]+\frac{{\rm d}}{{\rm d}t} G(t,r_1,r_2),
\end{equation}
which (minus the gauge function $G$) was found in \cite{TruF}.
Then, property (b) yields a Jacobi Last Multiplier for the
Gompertz's system (\ref{Gom}). The product of $M_{[r]}$ in
(\ref{GomMr}) with the Jacobian of the transformation
(\ref{Gomtr}) between $(w_1,w_2)$ and $(r_1,r_2)$ yields the
following  Jacobi Last Multiplier of system (\ref{Gom}), i.e.
\begin {equation}
M_{[w]}
=M_{[r]}\frac{\partial(r_1,r_2)}{\partial(w_1,w_2)}=\exp[-(a+A)t]\left
|
\begin {array} {cc}
{\displaystyle{\frac{1}{w_1}}}& 0\\ [0.3cm] 0 &
{\displaystyle{\frac{1}{w_2}}}
\end {array}\right|=\exp[-(a+A)t]\frac{1}{w_1w_2}, \label {GomM}
\end {equation}
and consequently a Lagrangian of the original system (\ref{Gom})
\begin{eqnarray}
L_{[w]}&=&\exp[-(a+A)t]\left[\log(w_1)\frac{\dot
w_2}{w_2}-\log(w_2)\frac{\dot w_1}{w_1}
-2a\log\left(\frac{w_2}{m_2}\right)\log(w_1) +2Bw_2
\right.\nonumber\\&&
           \left.
            -
2bw_1+2A\log\left(\frac{w_1}{m_1}\right)\log(w_2) - (A -
a)\log(w_1)\log(w_2)\right]
            +\frac{{\rm d}}{{\rm d}t} G(t,w_1,w_2).
\end{eqnarray}
This Lagrangian was not obtained in \cite{TruF}.

 We can transform
system (\ref{Gomt}) into an equivalent second-order ordinary
differential equation by eliminating, say, $r_2$. In fact from the
second equation in (\ref{Gomt}) one gets
\begin {equation}r_2=\log\left(\frac{\dot r_1 - A r_1}{B m_2}\right),\label{Gomr2}\end{equation}
and the equivalent second-order equation in $r_2$ is the following
\begin {equation}
\ddot r_1= \left(b m_1\exp(r_1)+a\log\left(\frac{\dot r_1 - A
r_1}{B m_2}\right)\right)(\dot r_1 - A r_1)+A\dot r_1.
 \label{Gomr1}
\end {equation}
 Using property (b) a Jacobi Last Multiplier for this equation
 can be  obtained. In fact we have to calculate
the Jacobian of the transformation between $(r_1,r_2)$ and
$(r_1,\dot r_1)$, namely (\ref{Gomr2}) and this yields a Jacobi
Last Multiplier of equation (\ref{Gomr1}), i.e.\footnote{Of
course, we do not consider any multiplicative constants because
they are  inessential.}
\begin {equation}
M_1 =M_{[r]}\frac{\partial(r_1,r_2)}{\partial(r_1,\dot
r_1)}=\exp[-(a+A)t]\frac{1}{\dot r_1-Ar_1}. \label {Gomr1M}
\end {equation}
 Then a
Lagrangian  can be   obtained by a double integration as in
(\ref{lagrint}), i.e.
\begin {equation}
L_1= \exp[-(a+A)t]\Big( ( \dot r_1-Ar_1)\log(\dot r_1-Ar_1)+ m_1
b\exp(r_1) - a r_1\log(B m_2) - a r_1\Big) +\frac{{\rm d}}{{\rm d}t} F(t,r_1).
\label{GomL1}
\end{equation}
The same Lagrangian (minus the gauge function $F$) was obtained in
\cite{TruF}.

\subsection{ Verhulst's model}
The Verhulst's model considered in \cite{TruF} is the following:
\begin{eqnarray} \dot w_1&=& w_1(A +B w_1+ f_1 w_2) \nonumber\\
 \dot w_2 &=& w_2(a + bw_2 + f_2w_1) \label{Ver}.
\end{eqnarray}
In order to derive a Jacobi Last Multiplier for this system from
(\ref{Meqi}), i.e.
\begin{equation}
\frac{{\rm d}}{{\rm d}t}\log \left(M_{[w]}\right) + (2B+ f_2)w_1 +
(2b+f_1)w_2  + a + A=0\label{VereqMw}
\end{equation}
we assume that $M_{[w]}$ has the following form:
\begin{equation}
M_{[w]}=w_1^{b_1}w_2^{b_2}\exp(b_3t), \label{VerhMw}
\end{equation}
where $b_i, (i=1,2,3)$ are constants to be determined. Replacing
this $M_{[w]}$ into (\ref{VereqMw}) yields
\begin{eqnarray}
b_1&=&\frac{ - 2B b+ b f_2 + f_1 f_2}{B b - f_1 f_2}\\
 b_2&=&\frac{- 2 B b + B f_1 + f_1 f_2}{B b - f_1 f_2}\\
 b_3&=&\frac{A B b-A b f_2+a B b-a B f_1}{B b-f_1 f_2},
 \label{Verai}
\end{eqnarray}
if $B b-f_1 f_2\neq 0$, and therefore if no condition is imposed
on the parameters in Verhulst's model. Consequently a Lagrangian of system (\ref{Ver})
 is
\begin{eqnarray}
L_{[w]}&=&\exp(b_3t)\left(w_2^{b_2}w_1^{b_1+1}\frac{\dot w_2}{b_1+1}
-w_2^{b_2+1}w_1^{b_1}\frac{\dot w_1}{b_2+1}\right.\nonumber \\&&\left.- w_2^{b_2+1}w_1^{b_1+1}
\left(2\frac{f_2w_1}{b_1 + 2}+2\frac{bw_2}{b_1 + 1}
   +\frac{2a(b_2 + 1) + b_3}{(b_1 + 1)(b_2 + 1)}\right)\right)+\frac{{\rm d}}{{\rm d}t} G(t,w_1,w_2)
\end{eqnarray}
that was not obtained in \cite{TruF}.

 We follow \cite{TruF} and introduce the
change of variables\footnote{It is not clear the reason of this
change of variables that was performed in \cite{TruF}.}
\begin{equation}
w_1=\exp(r_1),\quad\quad w_2=\exp(r_2)\label{Verr}
\end{equation}
and then system (\ref{Ver})  becomes
\begin{eqnarray} \dot r_1&=& A + B \exp(r_1)
+ f_1 \exp(r_2) \nonumber\\
 \dot r_2 &=& a+b \exp(r_2)+f_2 \exp(r_1)\label{Vert}.
\end{eqnarray}
We can transform this system into an equivalent second-order
ordinary differential equation by eliminating, say, $r_2$. In fact
from the second equation in (\ref{Vert}) one gets
\begin {equation}r_2=\log\left(\frac{ \dot r_1- B\exp(r_1) - A}{f_1}\right),\label{Verr2}
\end{equation}
and the equivalent second-order equation in $r_1$ is the following
\begin {eqnarray}
\ddot r_1&=&\frac{1}{f_1} \Big[\left(af_1 + b\dot r_1\right)\dot
r_1 + A^2 b +B \exp(2r_1)(B b - f_1f_2) - A (af_1 + 2b\dot r_1)
\nonumber
\\&& \quad \quad  - \exp(r_1)\Big(f_1\left( a B - f_2\dot r_1\right) +B (2b - f_1)\dot r_1 - A(2b B
- f_1f_2)\Big)\Big]. \label{Verr1}
\end {eqnarray}

Using property (b) a Jacobi Last Multiplier for this equation
 can be  obtained. In fact we have to calculate
the Jacobian of the transformation between $(w_1,w_2)$ and
$(r_1,\dot r_1)$,  and this yields a Jacobi Last Multiplier of
equation (\ref{Verr1}), i.e.\footnote{Of course, we do not
consider any multiplicative constants because they are
inessential.}
\begin {equation}
M_1 =M_{[w]}\frac{\partial(w_1,w_2)}{\partial(r_1,\dot r_1)}=
\exp(b_1r_1 + b_3t)[\dot r_1-A-B\exp(r_1)]^{b_2}(b_2+2)(b_2+1).
\label {VerM1}
\end {equation}
    Then a Lagrangian  can be   obtained by a double integration as in
(\ref{lagrint}), i.e.
\begin {equation}
L_1= \exp(b_1r_1)\exp(b_3 t)[\dot r_1-A-B\exp(r_1)]^{b_2+2}+\frac{{\rm d}}{{\rm d}t}
F(t,r_1). \label{VerL1}
\end{equation}
The same Lagrangian (minus the gauge function $F$) was obtained in
\cite{TruF}.\\
Since a Jacobi Last Multiplier of system (\ref{Vert}) is
\begin{equation}
M_{[r]} =M_{[w]}\frac{\partial(w_1,w_2)}{\partial(r_1,r_2)}=
\exp[(b_1+1)r_1+(b_2+1)r_2 + b_3t],
\end{equation}
analogously a Lagrangian of system (\ref{Vert}) is
\begin{eqnarray}
L_{[r]}&=&\exp((b_1+1)r_1+(b_2+1)r_2+b_3t)\left(\frac{\dot r_2}{b_1+1}
-\frac{\dot r_1}{b_2+1}\right.\nonumber \\
&&\left.-
\left(2\frac{f_2\exp(r_1)}{b_1 + 2}+2\frac{b\exp(r_2)}{b_1 + 1}
   +\frac{2a(b_2 + 1) + b_3}{(b_1 + 1)(b_2 + 1)}\right)\right)+\frac{{\rm d}}{{\rm d}t} G(t,r_1,r_2).
\end{eqnarray}
This Lagrangian (minus the gauge function $G$) was also obtained in
\cite{TruF}.\\
\subsection{ Host-Parasite model}
As stated in \cite{TruF}, ``a simple mathematical model which
describes the interaction between a host and its parasite and
which takes into account the non-linear effects of the host
population size on the growth rate of the parasite population is
given by the equations \cite{LesG}"
\begin{eqnarray} \dot w_1&=& (a-b w_2)w_1 \nonumber\\
 \dot w_2 &=& \left(A-B\frac{w_2}{w_1}\right)w_2.  \label{HP}
\end{eqnarray}
As in the previous example it is easy to derive that a Jacobi Last
Multiplier is
\begin {equation} M_{[w]}=\frac{\exp(At)}{w_1 w_2^2}, \label{HPMs}
\end{equation}
and consequently a Lagrangian of system (\ref{HPt})
\begin{eqnarray}
L_{[w]}&=&\exp[At]\left[\log(w_1)\frac{\dot w_2}{w_2^2}+\frac{\dot
w_1}{w_1w_2}-2\frac{a}{w_2} - 2\frac{B}{w_1}\right.\nonumber\\&&
           \left. -\log(w_1)\frac{A}{w_2}-2b\log(w_2)\right]+\frac{{\rm d}}{{\rm d}t} G(t,w_1,w_2).
\end{eqnarray}
This Lagrangian was obtained in \cite{TruF}.\\
In order to simplify system (\ref{HP})  we introduce the change of
variables\footnote{This change of variables was not performed in
\cite{TruF}.}
\begin{equation}
w_1=r_1\exp(at),\quad\quad w_2=r_2\exp(At)\label{HPtr}
\end{equation}
and then system (\ref{HP})  becomes
\begin{eqnarray} \dot r_1&=& - b\exp(At)r_1r_2\nonumber\\
 \dot r_2 &=& - \frac{B\exp(At)r_2^2}{\exp(at)r_1} \label{HPt}.
\end{eqnarray}
Since a Jacobi Last Multiplier of system (\ref{HPt}) is
\begin{equation}
M_{[r]} =M_{[w]}\frac{\partial(w_1,w_2)}{\partial(r_1,r_2)}=
\frac{1}{r_1r_2^2},
\end{equation}
analogously a Lagrangian of system (\ref{HPt}) is
\begin{eqnarray}
L_{[r]}&=&\frac{\log(r_1)\dot r_2}{r_2^2}
+\frac{\dot r_1}{r_1r_2}-2\exp(At)\frac{br_1\log(r_2)\exp(at)+B}{r_1\exp(at)}
+\frac{{\rm d}}{{\rm d}t} G(t,r_1,r_2).
\end{eqnarray}
This Lagrangian  was obviously not obtained in
\cite{TruF}.\\
 We can transform system
(\ref{HPt}) into an equivalent second-order ordinary differential
equation by eliminating, say, $r_2$. In fact from the first
equation in (\ref{HPt}) one gets
\begin {equation}r_2= -\frac{\dot r_1}{b\exp(At)r_1},\label{HPr2}
\end{equation}
and the equivalent second-order equation in $r_1$ is the following
\begin {equation}
\ddot r_1=\frac{b\exp(at)r_1 + B}{ b\exp(at)r_1^2}\,\dot r_1^2 +
A\dot r_1. \label{HPeq}
\end {equation}
Using property (b) a Jacobi Last Multiplier for this equation
 can be  obtained. In fact we have to calculate
the Jacobian of the transformation between $(r_1,r_2)$ and
$(r_1,\dot r_1)$,  and this yields a Jacobi Last Multiplier of
equation (\ref{HPeq}), i.e.\footnote{Of course, we do not consider
any multiplicative constants because they are inessential.}
\begin {equation}
M_1 = M_{[r]}\frac{\partial(r_1,r_2)}{\partial(r_1,\dot
r_1)}=\frac{b^2\exp(2At)r_1}{\dot r_1^2}\left |
\begin {array} {cc}
1& 0\\ [0.3cm] {\displaystyle{\frac{\dot r_1}{b\exp(At)r_1^2}}} &
-{\displaystyle{\frac{1}{b\exp(At)r_1}}}
\end {array}\right|=-
\frac{b\exp(At)}{\dot r_1^2}. \label {HPM1}
\end {equation}
 Then a
Lagrangian  can be   obtained by a double integration as in
(\ref{lagrint}), i.e.
\begin {equation}
L_1= b\exp(At)\log(\dot r_1)- b\exp(At)\log(r_1) +
\frac{B\exp(At)}{\exp(at)r_1}+\frac{{\rm d}}{{\rm d}t} F(t,r_1). \label{HPL1}
\end{equation} This Lagrangian was not obtained in \cite{TruF}.

\section{Final remarks}
As stated by Paine  ``Among the mathematical results obtained by
studying the inverse problem of mechanics, it is the explicit
algorithms for constructing Lagrangians that offer the model
builder the most practical benefit." (omissis) ``This gives one
hope of finding an integral or constant of motion for the
dynamical system of interest without the hardship of solving the
system of equations." \cite{Paine}.

This paper deals with Jacobi Last Multiplier and its connection to
the inverse problem
 of calculus of variation, namely finding one or more Lagrangians for either systems of two first-order equations or
single second-order equations.

We could not find better words than those used by Trubatch and
Franco at the end of their paper ``These results clearly show that
as the interaction between the populations becomes more complex
the corresponding Lagrangian becomes more complicated and
difficult to find" \cite{TruF}. We hope to address this problem in a future publication.

\section*{Acknowledgements}
This work was undertaken while K.M.T. was visiting Professor M.C. Nucci and the Dipartimento di
Matematica e Informatica, Universit\`a di Perugia. K.M.T.
gratefully acknowledges the support of the Italian Istituto
Nazionale Di Alta Matematica ``F. Severi'' (INDAM), Gruppo
Nazionale per la Fisica Matematica (GNFM), Programma Professori
Visitatori.

 \end{document}